# Tracking Air Pollution in China: Near Real-Time PM$_{2.5}$ Retrievals from Multiple Data Sources

*Guannan Geng[1,2], Qingyang Xiao[1], Shigan Liu[3], Xiaodong Liu[1], Jing Cheng[3], Yixuan Zheng[4], Dan Tong[3], Bo Zheng[5], Yiran Peng[3], Xiaomeng Huang[3], Kebin He[1,2] and Qiang Zhang[3]*

[1]State Key Joint Laboratory of Environment Simulation and Pollution Control, School of Environment, Tsinghua University, Beijing 100084, China

[2]State Environmental Protection Key Laboratory of Sources and Control of Air Pollution Complex, Beijing 100084, China

[3]Ministry of Education Key Laboratory for Earth System Modelling, Department of Earth System Science, Tsinghua University, Beijing 100084, China

[4]Center of Air Quality Simulation and System Analysis, Chinese Academy of Environmental Planning, Beijing 100012, China

[5]Institute of Environment and Ecology, Tsinghua Shenzhen International Graduate School, Tsinghua University, Shenzhen 518055, China

*Correspondence to*: Qiang Zhang (qiangzhang@tsinghua.edu.cn)

**Abstract.** Air pollution has altered the Earth's radiation balance, disturbed the ecosystem and increased human morbidity and mortality. Accordingly, a full-coverage high-resolution air pollutant dataset with timely updates and historical long-term records is essential to support both research and environmental management. Here, for the first time, we develop a near real-time air pollutant database known as Tracking Air Pollution in China (TAP, http://tapdata.org/) that combines information from multiple data sources, including ground measurements, satellite retrievals, dynamically updated emission inventories, operational chemical transport model simulations and other ancillary data. Daily full-coverage $PM_{2.5}$ data at a spatial resolution of 10 km is our first near real-time product. The TAP $PM_{2.5}$ is estimated based on a two-stage machine learning model coupled with the synthetic minority oversampling technique and a tree-based gap-filling method. Our model has an averaged out-of-bag cross-validation $R^2$ of 0.83 for different years, which is comparable to those of other studies, but improves its performance at high pollution levels and fills the gaps in missing AOD on daily scale. The full coverage and near real-time updates of the daily $PM_{2.5}$ data allow us to track the day-to-day variations in $PM_{2.5}$ concentrations over China in a timely manner. The long-term records of $PM_{2.5}$ data since 2000 will also support policy assessments and health impact studies. The TAP $PM_{2.5}$ data are publicly available through our website for sharing with the research and policy communities.

# 1 Introduction

With rapid urbanization and economic growth, anthropogenic emissions of reactive gases, aerosols and aerosol precursors are being emitted into the atmosphere, and these substances substantially changed the atmospheric composition. Consequently, worsening air pollution has altered the Earth's radiation balance, distressed the ecosystem and increased the risks of human morbidity and mortality[1,2]. In particular, one of the major air pollutants in China is fine particulate matter ($PM_{2.5}$), which could cause serious health problems[3,4] and reduce visibility[5]. Hence, understanding the spatial and temporal variations of the ambient $PM_{2.5}$ concentration constitutes the basis for research studies associated with air pollution, climate change and environmental health. It follows, then, that a complete-coverage high-resolution $PM_{2.5}$ dataset with timely updates and historical long-term records is essential to support both scientific research and environmental management. A complete-coverage dataset would allow us to obtain a full spatial picture of $PM_{2.5}$ pollution and avoid exposure misclassification in epidemiological studies[6,7]. Moreover, real-time or near real-time updates of $PM_{2.5}$ data would help us track substantial changes in air pollution during haze events or special times such as the coronavirus pandemic. In addition, a historical long-term dataset benefitting from a consistent methodology could support clean air policy assessments and health impact studies[8-11].

Several data sources could provide information about $PM_{2.5}$ pollution. Among them, ground measurements are the most accurate way to obtain ambient $PM_{2.5}$ concentrations. However, due to the installation and maintenance costs of ground networks, monitoring stations are usually sparse and unevenly distributed, with most of the stations located in urban areas[11,12]. Moreover, $PM_{2.5}$ ground networks in China were established in 2013, so prior data are unavailable. As an alternative, chemical transport models (CTMs) could provide complete-coverage simulations of $PM_{2.5}$ concentrations and could reproduce the spatial and temporal trends of $PM_{2.5}$ concentrations when using reasonable emission inventories[10, 13]. However, biases still exist in the simulated absolute values of $PM_{2.5}$ due to uncertainties in emission inventories[14] and the lack of certain physical and chemical processes in the model[15-17]. Satellite-retrieved aerosol optical depth (AOD) data, which can reflect the aerosol abundance in the atmosphere, have the advantage of long-term records and a high resolution. However, AOD data are missing during haze events, on cloudy days, and over bright surfaces such as desert and areas covered with snow, and the spatial distribution of missing data is sometimes nonrandom, which might cause biases in the long-term average[18]. Accordingly, a dataset that combines data from multiple sources is needed to take advantage of all available information and to meet the requirements of such a dataset, namely, a high accuracy, a full spatial coverage, a long temporal span, and real-time updates.

Previous studies have developed different methods to fuse two or more of the above datasets with other ancillary data in China to improve the estimation of $PM_{2.5}$[19-28]. These methods include CTM-based algorithms[26, 27], physical models[21], statistical models such as linear mixed-effects models and generalized additive models[19, 22], and machine learning models such as random forest and extreme gradient boosting[20, 23-25, 28]. As a result, numerous researchers have developed historical datasets in China, e.g., 10-km $PM_{2.5}$ data between 2000 and 2016 by Xue et al.[25] and 1-km $PM_{2.5}$ data between 2000 and 2018 by Wei et al.[23]. However, only some of these studies fill the gaps in AOD data using CTM simulations on a daily scale and achieve full-coverage daily $PM_{2.5}$ concentrations[24, 25, 28]. And previous studies usually underestimate $PM_{2.5}$ concentrations on highly polluted days (e.g., $PM_{2.5}>150$ μg/m$^3$) due to the small sample size of high-pollution cases and highly nonlinear relationship between $PM_{2.5}$ and AOD[25, 29]. Furthermore, none of these works provide near real-time $PM_{2.5}$ data publicly. Consequently, a near real-time dataset with gap-filled daily PM$_{2.5}$ estimates in China that can be shared with the research and policy communities remains lacking.

In this study, we develop the Tracking Air Pollution in China (TAP, http://tapdata.org/) database based on an operational CTM, a two-stage machine learning model and a gap-filling method[30]. Our goal is to combine information from multiple data sources and provide near real-time (i.e., one-day delay) PM$_{2.5}$ data on a daily scale with complete coverage at a spatial resolution of 10 km since 2000 to support related studies and environmental management. TAP is fully integrated on the cloud-computing platform, which allows users to conveniently access all customized data products online. Due to the downloading, processing and modeling procedures, the PM$_{2.5}$ data of the previous day are available to the public at approximately 10:00 local time.

## 2 Data and methods

### 2.1 Multisource input data

Table 1 summarizes all the input data used in this study, including ground measurements, satellite AOD, input data for the Weather Research and Forecasting/Community Multiscale Air Quality Modeling System (WRF/CMAQ) and other ancillary data such as meteorological fields, land use data, population, and elevation. The PM$_{2.5}$ measurements updated every hour are collected from the national air quality monitoring network (http://www.cnemc.cn/), which includes ~1,600 stations over China. Continuous identical data over three hours are excluded, and then the daily mean PM$_{2.5}$ is calculated only if at least 12 hourly measurements are available. The PM$_{2.5}$ measurements are matched to the 10-km grid cells they fall in.

Moderate Resolution Imaging Spectroradiometer (MODIS) Collection 6 level 2 aerosol products[31] from both Aqua and Terra at a spatial resolution of 0.1° are downloaded from the National

Aeronautics and Space Administration (NASA, https://ladsweb.modaps.eosdis.nasa.gov/). We use AOD measurements retrieved by both the Dark Target (DT) algorithm[31] and the Deep Blue (DB) algorithm[32] from Terra and Aqua to improve the spatial coverage of AOD data. A daily linear regression is first fitted between the DT and DB AOD when both are available for Terra and Aqua separately and then used to fill the missing AOD when only the DT AOD or the DB AOD is valid. Then, the AOD averaged between the DT and DB AOD is calculated for Terra and Aqua separately. Similarly, a second linear regression is fitted between the Terra and Aqua AOD to fill the missing AOD when only one of them is available. The average Terra and Aqua AOD is used to represent the daily aerosol loading[33].

The WRF/CMAQ modeling system is included in our work to provide daily $PM_{2.5}$ simulations. We employ the National Center for Environmental Prediction Final Analysis (NCEP-FNL, https://rda.ucar.edu/datasets/ds083.2/) and the Global Forecast System (NCEP-GFS, https://www.nco.ncep.noaa.gov/pmb/products/gfs/) to drive the WRF model. Anthropogenic emissions are taken from the Multi-resolution Emission Inventory in China (MEIC, http://meicmodel.org/)[34,35], which is updated in a timely manner using a bottom-up approach based on near real-time activity indicators[36]. More details about the dynamic emissions can be found in Zheng et al.[36], and a more in-depth description of the WRF/CMAQ model is provided in Sect. 2.2.1. Daily $PM_{2.5}$ simulations from the WRF/CMAQ model are interpolated into the 10-km grid using the inverse distance weighting (IDW) method.

The meteorological analysis data are taken from the Modern-Era Retrospective Analysis for Research and Applications Version 2 (MERRA-2) dataset at a resolution of 0.5° × 0.625° for the period 2000–2013[37] and from the Goddard Earth Observing System Forward Processing (GEOS-FP) dataset at a resolution of 0.25° × 0.3125° for the period ranging from 2013 to the present[38].

We utilize the following parameters extracted from the analysis data: surface albedo, surface pressure, surface incoming shortwave flux, surface net downward shortwave flux, surface net downward longwave flux, total incoming shortwave flux, total net downward shortwave flux, total latent energy flux, cloud area fraction for low clouds, total cloud area fraction, total column ozone, total column odd oxygen, bias-corrected total precipitation, total precipitable ice water, total precipitable water vapor, total precipitable liquid water, evaporation from turbulence, 2-meter specific humidity, 2-meter dew point temperature, 2-meter air temperature, 2-meter eastward wind (U wind), 2-meter northward wind (V wind), 10-meter U wind, 10-meter V wind, 10-meter wind speed, U wind at 500 hPa, V wind at 500 hPa, U wind at 850 hPa, V wind at 850 hPa, planetary boundary layer height and snowfall. Daily averaged meteorological analysis data are interpolated into the 10-km grid using the IDW approach.

We also download the urban and rural land cover classification data at a spatial resolution of 30 m from Gong et al.[39] (http://data.ess.tsinghua.edu.cn/) and elevation data at a spatial resolution of 30 m from the Global Digital Elevation Model (GDEM) version 2 (https://earthexplorer.usgs.gov/). Population data at a spatial resolution of 1 km are taken from the Gridded Population of the World (GPW) version 4 dataset and are calibrated using the WorldPop dataset at the county level and the total population reported in China City Yearbooks. The land cover data and elevation data are averaged within the 10-km grid while the population data are summed within the 10-km grid.

## 2.2 Algorithm description

### 2.2.1 Operational WRF/CMAQ modeling system

The TAP database includes an operational WRF/CMAQ modeling system to provide daily $PM_{2.5}$ simulations as one of the input data sources, which could improve the accuracy of $PM_{2.5}$ estimations and fill the gaps caused by missing AOD data. The WRF model version 3.9.1 and CMAQ model version 5.2 (https://www.cmascenter.org/cmaq/) are used in our work. The simulation domain covers all of China with a horizontal resolution of 36 km. The vertical resolution is designed as 46 sigma levels from the ground surface to 100 hPa for WRF but only 28 vertical layers in CMAQ after the processing of the Meteorology-Chemistry Interface Processor (MCIP). For the WRF model, the NCEP-FNL and NCEP-GFS data are used to provide the initial and boundary conditions, while the NCEP-GFS sea surface temperature (SST) reanalysis data and NCEP Automated Data Processing (ADP) global observational weather data are used for analysis, observation, and soil nudging. The parameterization scheme follows Cheng et al.[40] with the Kain-Fritsch cumulus physics scheme version 2[41] modified to the Grell-Freitas ensemble scheme[42]. For the CMAQ model, we use the CB05 gas-phase mechanism with the CMAQv5.1 update and sixth-generation CMAQ aerosol mechanism (AERO6). The chemical initial and boundary conditions are derived from ASCII vertical profile data.

The dynamically updated anthropogenic emissions for mainland China are taken from the MEIC[34-36]. Emissions for other Asian countries and regions are obtained from the MIX inventory[43]. Biogenic emissions are calculated by the Model of Emissions of Gases and Aerosols from Nature (MEGAN) version 2.1[44]. Sea salt and dust aerosol emissions are calculated online by the CMAQ model.

The simulated $PM_{2.5}$ concentrations from our WRF/CMAQ model have been fully evaluated against ground measurements in our previous studies[10, 40, 45]. Accordingly, the model performance statistics can meet the recommended performance criteria, and the simulated results have been used for policy assessment and health impact studies in China[10, 40, 45].

**2.2.2 Two-stage machine learning model**

A two-stage machine learning model coupled with the synthetic minority oversampling technique (SMOTE) developed in our previous study[46] is used to generate the TAP $PM_{2.5}$ data, as presented in Figure 1. In the first stage, we define a high-pollution indicator to improve the $PM_{2.5}$ estimations on highly polluted days, which are usually underestimated in statistical and machine learning models[23, 28]. This high-pollution indicator is calculated based on $PM_{2.5}$ observation data and describes whether the $PM_{2.5}$ observations at each location exceed the monthly mean by two standard deviations. As high-pollution events cover only 3.9% of our training dataset, which hinders the model's ability to characterize the associations between high-pollution events and other predictors, we adopt the SMOTE technique to resample our dataset and obtain a balance between high-pollution and normal samples. The resampled dataset is then used to train the first-stage random forest model with all the input data except for the CMAQ simulations, after which the predicted full-coverage high-pollution indicator is passed to the second-stage model as one of the input data. In the second stage, we use the residuals between the $PM_{2.5}$ measurements and the CMAQ $PM_{2.5}$ simulations as the dependent variable to train the second-stage random forest model. The predicted residuals combined with the CMAQ simulations represent the final $PM_{2.5}$ estimations.

Compared with the models presented in previous studies, our model has two major advantages. In the first stage, the SMOTE algorithm balances the uneven proportion of high-pollution and normal data, which could improve the model performance at high $PM_{2.5}$ levels. In the second stage, using the residuals between simulated and measured $PM_{2.5}$ enhances the variability of the dependent data, which could enhance the responses of predictors to $PM_{2.5}$ variations, thus improving the prediction accuracy. We design a sensitivity test model (*Sens*) without the SMOTE technique and using $PM_{2.5}$ measurements as the dependent variable to show our model improvements.

### 2.2.3 Gap-filling method

Our previous study[30] evaluated different gap-filling strategies and proposed a binary tree-based algorithm coupled with WRF/CMAQ simulations to fill the gaps in missing AOD. As the missingness of satellite AOD are primarily related to meteorological conditions (e.g., cloudy, rainy days) and $PM_{2.5}$ pollution (e.g., highly polluted days), the tree-based algorithm could directly predict missing $PM_{2.5}$ by mining the relationship between availability status of satellite data, $PM_{2.5}$ concentrations and other supporting information[47]. This method is robust at characterizing the spatial patterns of $PM_{2.5}$ without generating artificially oversmoothed $PM_{2.5}$ spatial distributions and is efficient for use in a near real-time data product[30]. In each step of our two-stage model, a dichotomous predictor defined by whether the satellite AOD is available is constructed as the cut point of the first layer of the decision tree. This predictor serves to build the associations between satellite AOD availability, $PM_{2.5}$ concentration, and other supportive information, such as WRF/CMAQ simulations and meteorological conditions, and helps to fill the gaps in the final $PM_{2.5}$ estimations.

### 2.3 Operational process of the TAP $PM_{2.5}$ data

Figure 1 shows the operational process for generating the near real-time $PM_{2.5}$ product in TAP, which includes three steps: data downloading, data processing and $PM_{2.5}$ modeling. Data from multiple sources (summarized in Table 1) are routinely downloaded to the cloud-computing platform every day once trey are available. As these data are at different temporal and spatial resolutions, they are processed to match the 10-km grid defined in our work, as described in Sect. 2.1.

Multiple $PM_{2.5}$ models are built to develop $PM_{2.5}$ data from 2000 to date. For years when ground $PM_{2.5}$ measurements are available (i.e., 2013–2020), individual models are developed for these years using input data within each year. For the hindcast of $PM_{2.5}$ prior to 2013 when ground measurements are absent, a model trained with dataset between 2013–2019 is developed and validated to provide robust hindcasting power. For the near real-time product since Jan 2021, the training dataset contains data from the year 2020 and is updated every day on a rolling basis to include the most recent input data. The two-stage random forest model is trained by the updated dataset every day, and then near real-time $PM_{2.5}$ data are generated and uploaded to our website.

## 3 Evaluation of model performance

The performance of our two-stage model is evaluated through three cross-validation (CV) experiments: out-of-bag CV, spatial CV and by-year CV. The out-of-bag CV is the most commonly used CV for the random forest models that compares the $PM_{2.5}$ measurements with the predictions of out-of-bag samples. Spatial CV evaluates the model's ability to make predictions at locations without monitors; all the monitoring stations are randomly divided into five subsets, and each time, the model is trained using data from four subsets and tested on the data from the remaining subset. Similarly, by-year CV evaluates the model's hindcast prediction ability, which sequentially selects one year of data for testing and trains the model with the data from the remaining years.

Table 2 shows the CV results of our two-stage random forest models at the daily level, including the $R^2$ and root mean square error (RMSE) values between the CV estimates and the ground measurements. The PM$_{2.5}$ predictions from the out-of-bag CV show good agreements with the observations, with R$^2$ of 0.80–0.88 and RMSE of 13.9−22.1 μg/m$^3$ for different years. The spatial CV R$^2$ value decreases by 0.05–0.11 when compared with the out-of-bag CV, indicating that unobserved spatial trends contribute to the PM$_{2.5}$ predictions. The model's hindcast performance further decreases in the by-year CV, with an R$^2$ of 0.58 and RMSE of 27.5 μg/m$^3$, reflecting a slight overfit in the hindcast of PM$_{2.5}$ in years prior to 2013.

Our model's performance is comparable to that of models presented in other studies on the basis of the R$^2$ and RMSE values shown in Table 2. The statistical or machine learning models at the 10-km grid on a daily scale have ten-fold CV R$^2$ values ranging between 0.79 and 0.80 in China[19, 22, 24], which are similar to our out-of-bag CV results (i.e., 0.83 on average). Models with a 1-km grid have higher R$^2$ values[23, 28], which might be partially explained by the correlations between PM$_{2.5}$ and the 1-km AOD being higher than those between PM$_{2.5}$ and the 10-km AOD, as well as the substantial increase in collocated AOD–PM$_{2.5}$ pairs at a 1-km resolution than at a 10-km resolution for a larger sample size[48].

**4 Illustration of TAP capabilities**

**4.1 Near real-time updates**

Our TAP PM$_{2.5}$ product is the first near real-time PM$_{2.5}$ database in China based on multisource data, including ground measurements, satellite AOD, high-resolution emission inventories (i.e., the MEIC inventory) and WRF/CMAQ simulations. Several factors support the timely update of PM$_{2.5}$ data. First, the dynamic updates of anthropogenic emissions in China by the MEIC and the high-performance computer at Tsinghua University facilitate the operational simulation of the WRF/CMAQ model, which is an important data source for PM$_{2.5}$ estimations, as has been evaluated in previous studies[25, 28]. Second, we choose the tree-based algorithm to fill the gaps in PM$_{2.5}$ concentrations, which is accurate and has reasonable speed. Other methods for filling in AOD gaps such as the multiple imputation method make use of more PM$_{2.5}$ observations in the training dataset[18]; however, such a method has a much lower computation speed, and we found similar performances between these two gap-filling methods in our previous work[30]. Finally, the cloud-computing platform makes it possible to develop the model online and allows users to conveniently access all the data products. The daily dataset of PM$_{2.5}$ from TAP can be found through our website in near real time.

### 4.2 Improved performance on polluted days

Our two-stage model coupled with the SMOTE technique improves the PM$_{2.5}$ estimations on highly polluted days. Compared to the sensitivity test model (Sens) without SMOTE and using PM$_{2.5}$ measurements directly as the dependent variable, the two-stage model has a similar $R^2$ but higher regression slope (0.97 vs 0.94) when evaluated against ground measurements. Figure 2 shows a detailed comparison between our two-stage model and the Sens model using year 2015 as an example. Usually, PM$_{2.5}$ concentrations are underestimated over polluted days but a little overestimated in clean days. After adopting our two-stage model, the mean biases over China decrease by 5.9 μg/m$^3$ (Figure 2a). We also present examples of the estimated daily variations in PM$_{2.5}$ concentrations from TAP and Sens and find that TAP has better ability in capturing the concentrations peaks on polluted days.

### 4.3 Full-coverage on daily scale

Figure 3 shows full-coverage daily maps of the TAP PM$_{2.5}$ as well as the estimations without gap-filling and ground observations during an example period, 26–28 Dec 2020. On these days, satellite

AOD data only cover 16%–24% of the grid cells in China; thus, many PM$_{2.5}$ concentration hotspots are missing. Such missingness might cause biases in the averaged concentrations as the missing are sometimes nonrandom. In summer, AOD data are usually missing over southern China due to rain and clouds, while in winter, AOD data over northern China are usually missing due to snow cover and haze[49]. The nonrandom distribution of missing AOD causes negative biases in the average PM$_{2.5}$ in the north and positive biases in the average PM$_{2.5}$ in the south[18, 26]. After gap-filling with supportive information from the WRF/CMAQ simulations and meteorological data, the daily maps of PM$_{2.5}$ become complete and can more accurately capture the day-to-day variations in PM$_{2.5}$. For example, the TAP PM$_{2.5}$ maps successfully capture the PM$_{2.5}$ changes across the North China Plain during the haze event that occurred on 26–28 Dec 2020 (Figure 3). PM$_{2.5}$ pollution started to rise on 26 Dec 2020, and high PM$_{2.5}$ levels were found in southern Hebei and northern Shandong. Then, the pollution expanded, and on 28 Dec 2020, Shandong, Henan and northern Anhui were covered by haze exceeding 180 μg/m$^3$. Such patterns could not be captured using the PM$_{2.5}$ data without gap filling.

**4.4 Historical PM$_{2.5}$ trends since 2000**

The TAP PM$_{2.5}$ database is also able to provide historical trends of PM$_{2.5}$ from 2000 to the present (Figure 4). Indeed, PM$_{2.5}$ estimates prior to 2013 have larger uncertainties, as there are no observation data to calibrate and evaluate our models. The by-year CV indicates that the model's hindcast ability has a smaller R$^2$ and larger RMSE than the out-of-bag CV. However, we use the year-by-year emission inventory from MEIC and the long-term CMAQ simulations as important input data to support the PM$_{2.5}$ estimates before 2013, thereby providing the best available knowledge of the spatial and temporal trends of PM$_{2.5}$ concentrations in history over China. Moreover, the long-term satellite AOD dataset also provides valuable observational evidence of aerosol changes since 2000. We believe that the long-term trend of $PM_{2.5}$ constrained by these two datasets is reliable.

Figure 4 shows the $PM_{2.5}$ trends since 2000 in China. The TAP data capture the $PM_{2.5}$ increase before 2006 (when there is no efficient emission control policy) and the sharp drop in $PM_{2.5}$ concentrations after 2013 (when strict control measures were implemented). The peak of the national population-weighted mean $PM_{2.5}$ concentrations occurred in 2006 (70.9 μg/m$^3$), the starting year of the Eleventh Five Year Plan (FYP, 2006–2010), when flue-gas desulfurization devices were installed in coal-fired power plants. After that, the increasing trend of $PM_{2.5}$ concentrations was reversed. Since 2013, strict clean air policies have been implemented, i.e., the Air Pollution Prevention and Control Action Plan (2013–2017) and the Blue Sky Protection Campaign (2018–2020). The Air Pollution Prevention and Control Action Plan reduced the annual population-weighted mean $PM_{2.5}$ from 65.6 μg/m$^3$ in 2013 to 46.1 μg/m$^3$ in 2017, and the Blue Sky Protection Campaign further reduced the $PM_{2.5}$ concentrations to 37.0 μg/m$^3$ in 2020.

## 5 Discussion

In this study, we develop the TAP $PM_{2.5}$ database that couples real-time ground observations, near real-time satellite data and meteorological reanalysis data, and operational simulations from the WRF/CMAQ modeling system to provide $PM_{2.5}$ concentration data that are updated in a timely manner. Based on a two-stage machine learning model and gap-filling method, TAP provides daily full-coverage $PM_{2.5}$ concentrations at a spatial resolution of 10 km in near real time. All the data are publicly available through our website for sharing with the community.

Our work is subject to some limitations. First, our near real-time $PM_{2.5}$ products rely on the near real-time updates of all the input data (except for the land use, population and elevation data, which have update frequencies of yearly or longer). Delays in any of these datasets would influence the updates of our $PM_{2.5}$ data. Second, although we believe that the long-term spatial and temporal patterns of $PM_{2.5}$ concentrations prior to 2013 are reliable due to the reasonableness of the input data, the uncertainties in $PM_{2.5}$ on a daily scale are still larger than the daily $PM_{2.5}$ estimates after 2013. Finally, previous studies have shown that using 1-km AOD estimates from the Multi-Angle Implementation of Atmospheric Correction (MAIAC) algorithm would improve the model performance, as a finer resolution would result in better correspondence between the AOD and $PM_{2.5}$[48]. However, building near real-time models at 1 km would cause exponential increases in the required computing resources and storage. Therefore, we choose the 10-km $PM_{2.5}$ data as our first step for the TAP database.

In the future, we will continue to improve our methods and provide more air pollutant species and finer spatial resolution data. Accordingly, we will build the TAP database into a near real-time database of multiple air pollutants at different spatial and temporal resolutions based on multiple data sources.

**Acknowledgments**

This work was supported by the National Natural Science Foundation of China (42005135, 42007189, 41921005, and 41625020).

**Table 1: Summary of the datasets used in this study from multiple sources.**

| Data category | Data name | Spatial resolution | Temporal frequency | Data source |
|---|---|---|---|---|
| Ground observations | $PM_{2.5}$ measurements | Point | Hourly | http://www.cnemc.cn/ |
| Satellite AOD | MODIS Terra AOD | ~10 km | Daily | https://ladsweb.modaps.eosdis.nasa.gov/ |
| | MODIS Aqua AOD | ~10 km | Daily | https://ladsweb.modaps.eosdis.nasa.gov/ |
| Input for WRF/CMAQ | NCEP/FNL | 1° | Daily | https://rda.ucar.edu/datasets/ds083.2/ |
| | NCEP/GFS | 1° | Daily | https://www.nco.ncep.noaa.gov/pmb/products/gfs/ |
| | MEIC emissions | 36 km | Monthly | http://meicmodel.org/ |
| Meteorological analysis data | MERRA-2 | 0.5°×0.625° | 3-hourly | https://gmao.gsfc.nasa.gov/reanalysis/MERRA-2/ |
| | GEOS-FP | 0.5°×0.625° | 6-hourly | https://gmao.gsfc.nasa.gov/GMAO_products/NRT_products.php |
| Land use data | FROM-GLC | 30 m | Yearly | http://data.ess.tsinghua.edu.cn/ |
| Population | GPW v4 | 1 km | Yearly | https://beta.sedac.ciesin.columbia.edu/ |
| | WorldPop | County | Yearly | https://www.worldpop.org/ |
| | China City Yearbooks | National | Yearly | |
| Elevation | GDEM | 30 m | - | https://earthexplorer.usgs.gov/ |

**Table 2: Model performance compared with other studies developing national PM$_{2.5}$ datasets in China.**

| | Gap-filled | Spatial resolution | Temporal resolution | CV type | CV R$^2$ | RMSE (μg/m$^3$) |
|---|---|---|---|---|---|---|
| Ma et al.[22] | No | 10 km | Daily (2004–2013) | Ten-fold CV | 0.79 | 27.4 |
| Fang et al.[19] | No | 10 km | Daily (2013–2014) | Ten-fold CV | 0.80 | 22.8 |
| He and Huang[50] | No | 3 km | Daily (2015) | Ten-fold CV | 0.80 | 18.0 |
| Xiao et al.[24] | Yes | 10 km | Daily (2013–2017) | Ten-fold CV | 0.79 | 21.0 |
| Xue et al.[25] | Yes | 10 km | Daily (2000–2016) | By-year CV | 0.61 | 27.8 |
| Liang et al.[20] | Yes | 1 km | Monthly (2000–2016) | Ten-fold CV | 0.93 | 6.2 |
| Wei et al.[23] | No | 1 km | Daily (2013–2018) Monthly (2000–2018) | Ten-fold CV | 0.86–0.90 | 10.0–18.4 |
| Huang et al.[28] | Yes | 1 km | Daily (2013–2019) | Ten-fold CV | 0.87−0.88 | 11.9–21.9 |
| TAP PM$_{2.5}$ | Yes | 10 km | Daily (2000–current) | By-year CV | 0.62 | 27.7 |
| | | | | Out-of-bag CV (individual years) | 0.80−0.88 | 13.9−22.1 |
| | | | | Spatial CV (individual years) | 0.69−0.83 | 14.6−26.4 |
| | | | | By-year CV (hindcast model) | 0.58 | 27.5 |

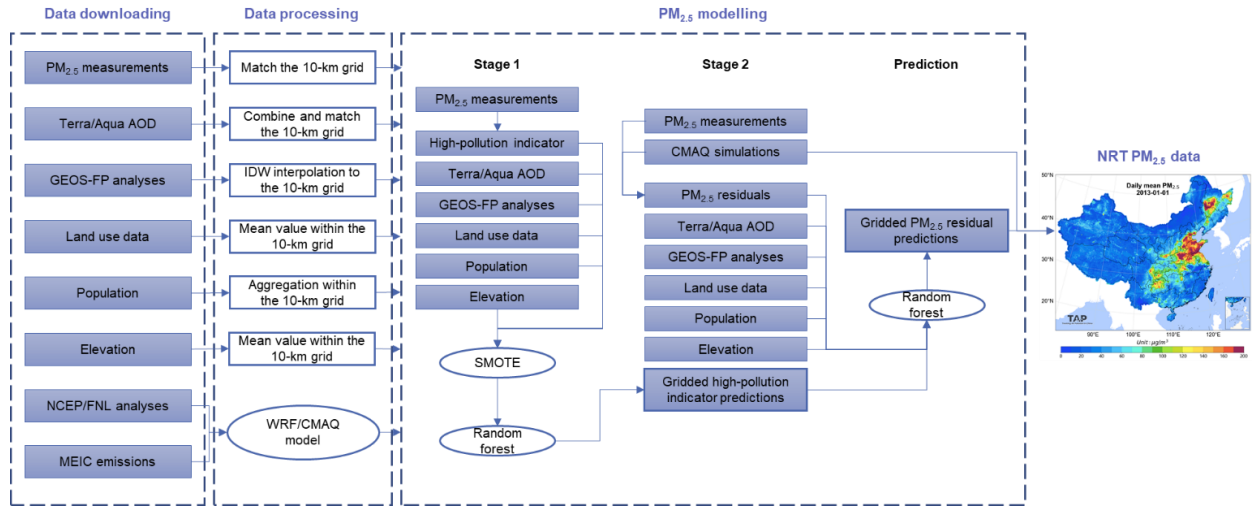

**Figure 1: Operational process of the near real-time PM$_{2.5}$ data generated from TAP.**

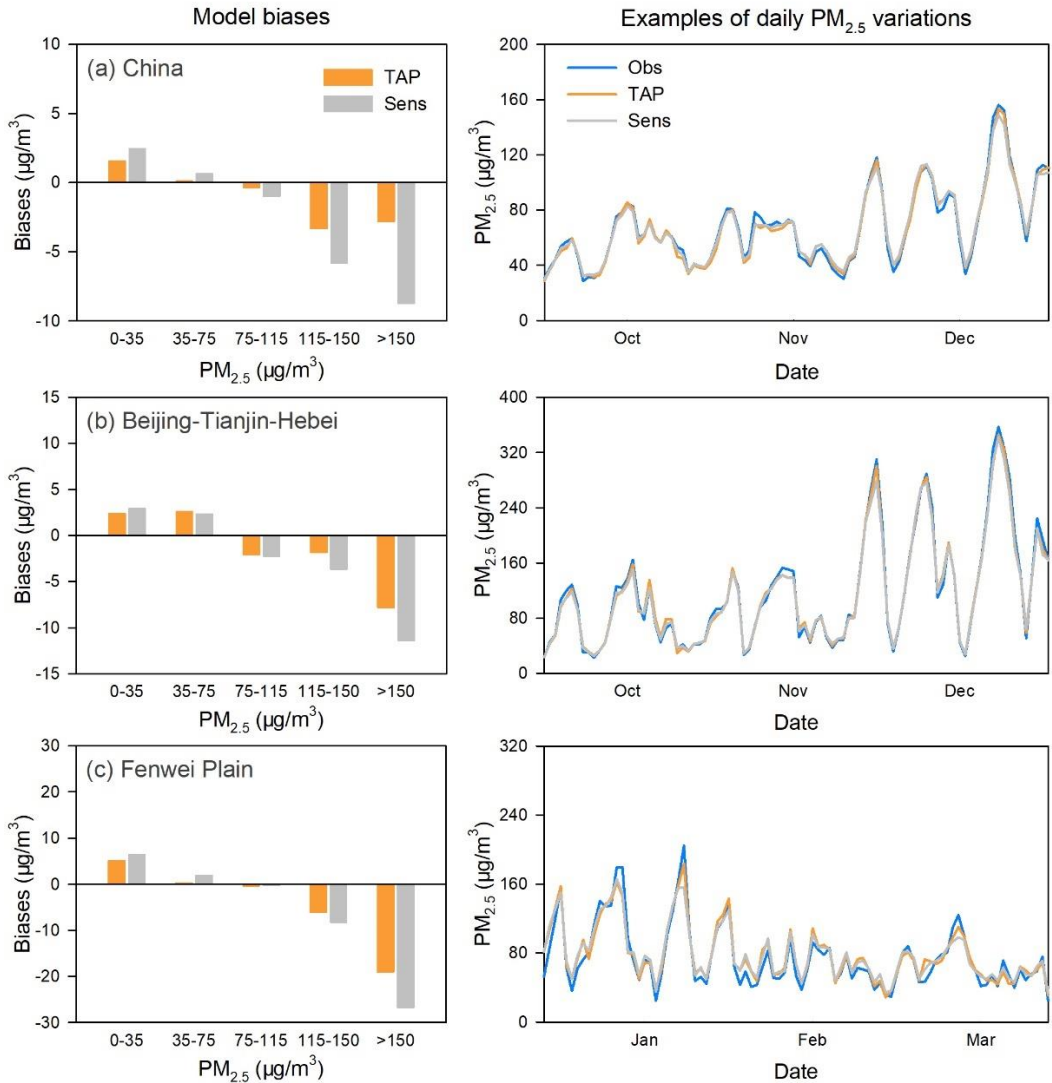

**Figure 2: Comparison between the two-stage model in TAP and the sensitivity test model Sens. Left column: PM$_{2.5}$ biases from TAP (orange) and Sens (grey) under different PM$_{2.5}$ pollution levels in (a) China, (b) the Beijing-Tianjin-Hebei region and (c) the Fenwei Plain. Right column: examples of daily PM$_{2.5}$ variations from ground observations (blue), TAP (orange) and Sens (grey).**

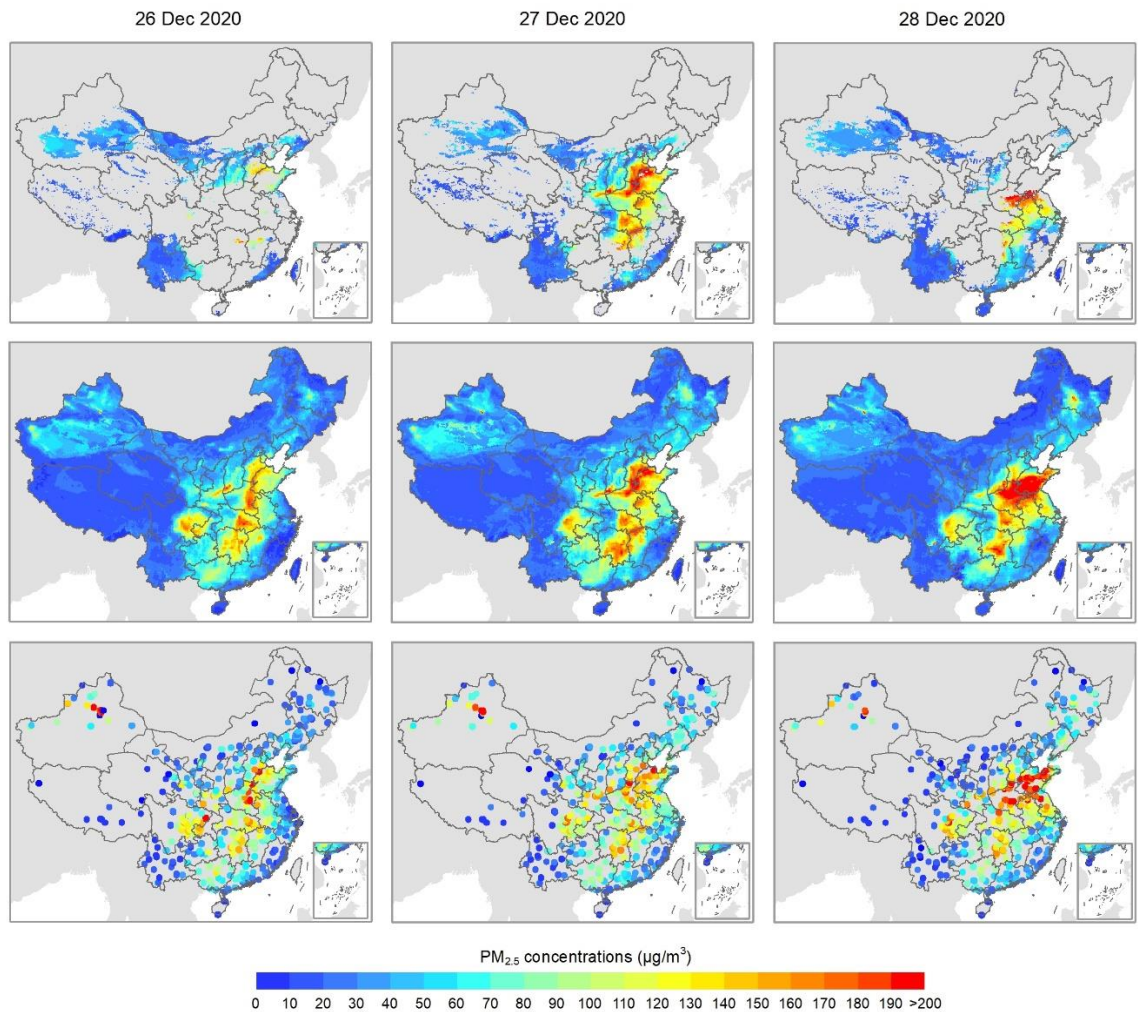

**Figure 3: Daily full-coverage PM$_{2.5}$ concentrations from TAP (middle row) compared with estimations without gap filling (top row) and ground observations (bottom row). Data for 26–28 Dec 2020 are shown as examples.**

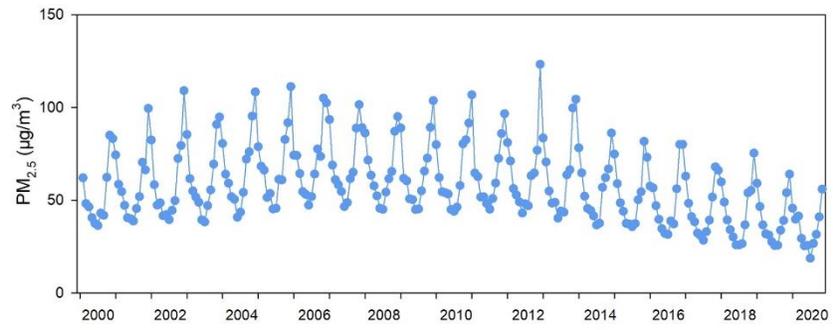

**Figure 4: Population-weighted mean PM$_{2.5}$ in China from 2000 to 2020 on a monthly scale.**